\documentclass[multphys,vecphys]{svmult}
\usepackage{makeidx} 
\ifx\pdfoutput\undefined
  \usepackage[dvips]{graphicx}
  \DeclareGraphicsExtensions{.eps}
\else
  \usepackage[pdftex]{graphicx}
  \DeclareGraphicsExtensions{.pdf}
\fi
\usepackage{multicol} 
\usepackage[bottom]{footmisc}

\makeindex  

\begin{document}

\title*{Baryonic loading and $e^+e^-$ rate equation in GRB sources}
\author{Remo Ruffini\inst{1,2}, Carlo Luciano Bianco\inst{1,2}, Gregory Vereshchagin\inst{1,2} \and She-Sheng Xue\inst{1,2}}
\authorrunning{Ruffini et al.}
\institute{ICRANet and ICRA, P.le della Repubblica 10, I--65100 Pescara, Italy.
\texttt{ruffini@icra.it}
\and Dip. Fisica, Univ. ``La Sapienza'', P.le A. Moro 5, I--00185 Roma, Italy.}

\maketitle

\begin{abstract}
The expansion of the electron-positron plasma in the GRB phenomenon is compared and contrasted in the treatments of M\'{e}sz\'{a}ros, Laguna and Rees, of Shemi, Piran and Narayan, and of Ruffini et al. The role of the correct numerical integration of the hydrodynamical equations, as well as of the rate equation for the electron-positron plasma loaded with a baryonic mass, are outlined and confronted for crucial differences.
\end{abstract}

\section{Introduction}

One of the earliest contributions in the theoretical understanding of GRBs is often referred to the work of \cite{Ruffini2-cr78}. Some crucial ideas which later became very important in this field were there presented (e.g. the relevance of an electron-positron plasma, the possibility of baryonic loading of such plasma, and the evolution from an optically thick to an optically thin phase). The treatment, based on qualitative consideration, provided little information about the dynamics. The electron-positron plasma was there purported to evolve toward a proliferation of photons and pairs, leading to a degradation of the mean energy of particles in the plasma and a consequent cooling. Such a plasma would become transparent on a very short time scale and no dynamical phases are envisaged making this model inappropriate for GRBs. The physical reasons of the non-validity of these crucial conclusions are being discussed elsewhere \cite{Ruffini2-RVA06}.

It soon became clear, however, following the work of \cite{Ruffini2-va} and \cite{Ruffini2-g86} that the major characteristic of a sudden energy release process in electron-positron plasma leads to a very rapid self acceleration of a shell of material, reaching ultra-relativistic regimes with Lorentz gamma factors in the range $10^2 < \gamma < 10^3$. The major results were obtained in \cite{Ruffini2-mlr92}, in \cite{Ruffini2-np}, in \cite{Ruffini2-bm95} and in \cite{Ruffini2-rswx99,Ruffini2-rswx00}.

In our model the analysis of the dynamical expansion of the electron-positron plasma is not just a topic of academic interest, it is indeed crucial to the description of the entire GRB phenomena. The initial process is the vacuum-polarization around the black hole \cite{Ruffini2-prx98}, followed by the dynamical expansion of the pair plasma, leading to an ultrarelativistic accelerated motion \cite{Ruffini2-rswx99,Ruffini2-rswx00}. Such acceleration leads to a shell of baryons with Lorentz gamma factor $\gamma \sim 100-300$ which, by interacting with the interstellar matter (ISM), gives origin to the afterglow. A first crucial signature in our model is carried by the radiation emitted at the moment of transparency, what we have called the Proper-GRB. The current models in the literature identify two different components in the long GRBs, the ``prompt radiation'' and the ``afterglow'', the first one being emitted by an ``inner engine'' (see e.g. \cite{Ruffini2-m06} and references therein). In our model, instead, the ``prompt radiation'' is an essential part of the ``afterglow'' and the entire long GRB phenomena is uniquely due to external shock processes.

For the above reasons we return, in this communication, to a comparison and contrast between our results and the ones in the current literature. Now we can address with clarity the analogies and the differences in the treatments and pin down the source of such differences as due to inaccurate theoretical work or / and equally inaccurate numerical computations.

\section{Contrasts in the dynamical description of the expanding plasma}

\begin{figure}
\centering
\includegraphics[width=0.495\hsize]{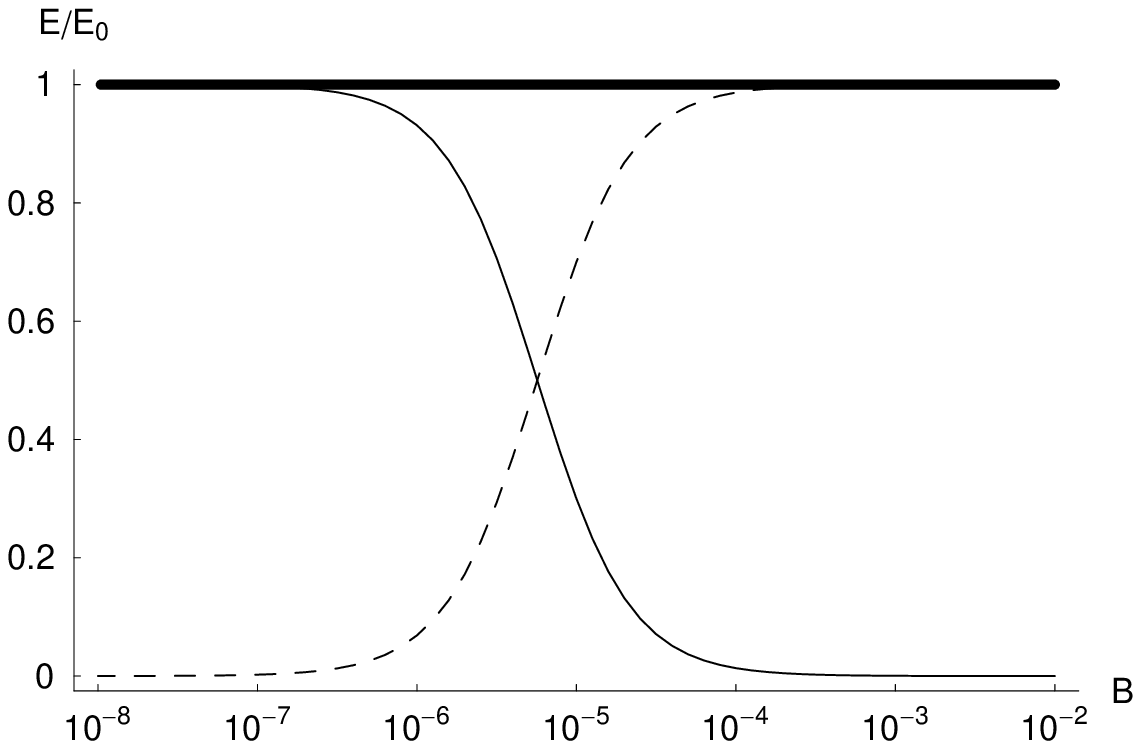}
\includegraphics[width=0.495\hsize]{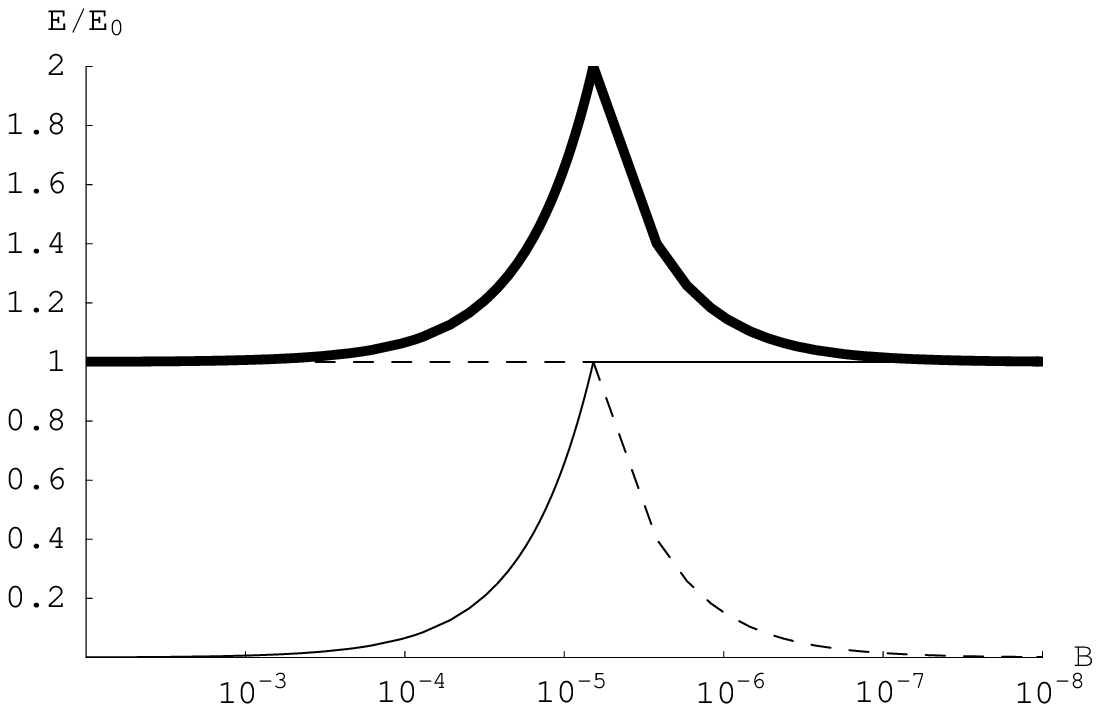}
\caption{{\bf Left:} Relative energy release in the form of photons emitted at transparency point $E_{\gamma}$ (solid line) and kinetic energy of the plasma $E_k$ (dashed line) of the baryons in terms of the initial energy $E_0$ of the electron-positron plasma as computed in our quasi-analytic model \cite{Ruffini2-rswx99,Ruffini2-rswx00}. The quantities are given as a function of the baryon loading parameter $B$. The bold line denotes the total energy of the system in terms of initial energy $E_0$ which is, as it should, constant and equal to $1$. {\bf Right:} The same quantities are computed in the M\'{e}sz\'{a}ros, Laguna and Rees model \cite{Ruffini2-mlr92}. Note that in Ref. \cite{Ruffini2-mlr92} the parameterization is done as a function of the dimensionless parameter $\eta=1/B$. The bold line denotes the total energy of the system in terms of initial energy $E_0$, which is not conserved in such a treatment \cite{Ruffini2-mlr92}.}
\label{Ruffini2-our}
\end{figure}

\begin{figure}
\centering
\includegraphics[width=0.495\hsize]{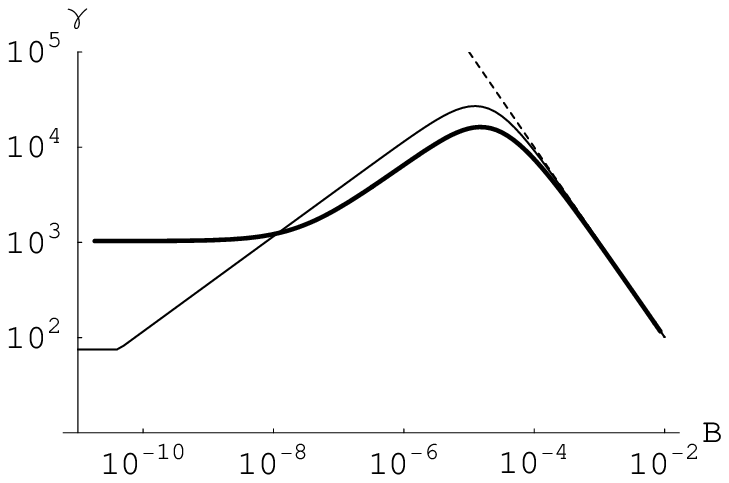}
\includegraphics[width=0.495\hsize]{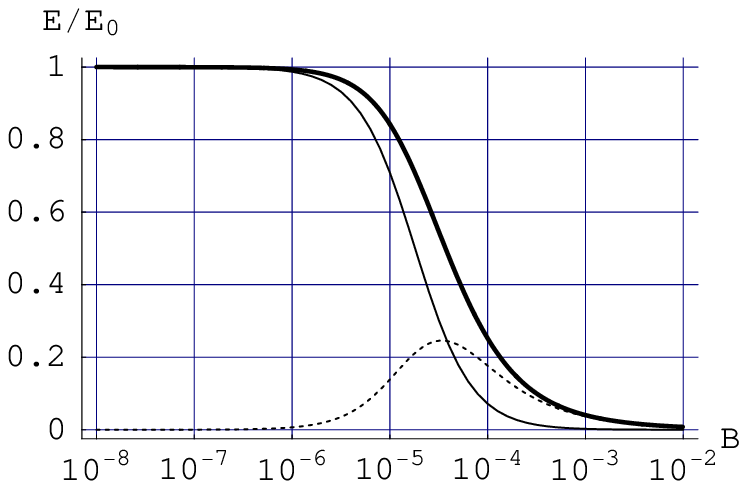}
\caption{{\bf Left:} Relativistic gamma factor when transparency is reached as a function of the baryon loading parameter $B$. The bold line denotes the numerical results obtained in Ref. \cite{Ruffini2-rswx99,Ruffini2-rswx00}, taking into due account the electron-positron rate equation. The plain line corresponds to the analytical estimate from Shemi and Piran model \cite{Ruffini2-sp}, neglecting the rate equation. The dashed line denotes the asymptotic value of the Lorentz gamma factor at transparency $\gamma_{asym}=B^{-1}$ (see Ref. \cite{Ruffini2-rswx00} for details). {\bf Right:} Relative energy release in the form of photons emitted at transparency point $E_{\gamma}$ of the GRB in terms of initial total energy $E_0$ depending on the baryon loading parameter $B$. The bold line represents our numerical results, already given in Fig. \ref{Ruffini2-our}-Left. The plain line shows the results for the analytic model of Shemi and Piran \cite{Ruffini2-sp}. The dashed line shows the difference between our numerical analysis, taking into proper account the rate equation \cite{Ruffini2-rswx99,Ruffini2-rswx00} and the approximate analytical \cite{Ruffini2-sp} results.}
\label{Ruffini2-npf}
\end{figure}

Our results are represented in Fig. \ref{Ruffini2-our}-Left, showing the energy release at transparency carried by photons and in the form of kinetic energy of the pulse, depending on the loading parameter $B=Mc^2/E_0$, where $M$ is the total mass of the plasma, $c$ is the speed of light, $E_0$ is initial energy of the system. In Fig. \ref{Ruffini2-our}-Right we represent the corresponding results by \cite{Ruffini2-mlr92} which are markedly different from ours, and do not fulfill the basic requirement of the conservation of energy. In Fig. \ref{Ruffini2-npf}-Left we represent the results of \cite{Ruffini2-sp} about the gamma factor at transparency which overlap with ours. The major difference between our treatment and the one by \cite{Ruffini2-np} and collaborators stems from different accuracy in the description of the electron-positron pairs equations which we have explicitly integrated in full details with the dynamical equations of the pulse. In Fig. \ref{Ruffini2-npf}-Right we represent some analogies and differences between the two treatments. In particular, the energy release in the form of photons at transparency, which is crucial for the detection of the P-GRB, is qualitatively correct in Ref. \cite{Ruffini2-sp} but underestimated in view of the simplified analytical approach adopted, which does not take into proper account the explicit integration of the electron-positron rate equation. The moment when the pulse reaches transparency and the corresponding radius are also different in simple analytical models and our detailed numerical computations, which leads to different predictions for the energetics.

In addition to the above differences, the treatment in \cite{Ruffini2-rswx00} clearly predicts an instability in the expanding plasma for a value $B > 10^{-2}$.

\section{Conclusions}

\begin{figure}
\centering
\includegraphics[width=\hsize]{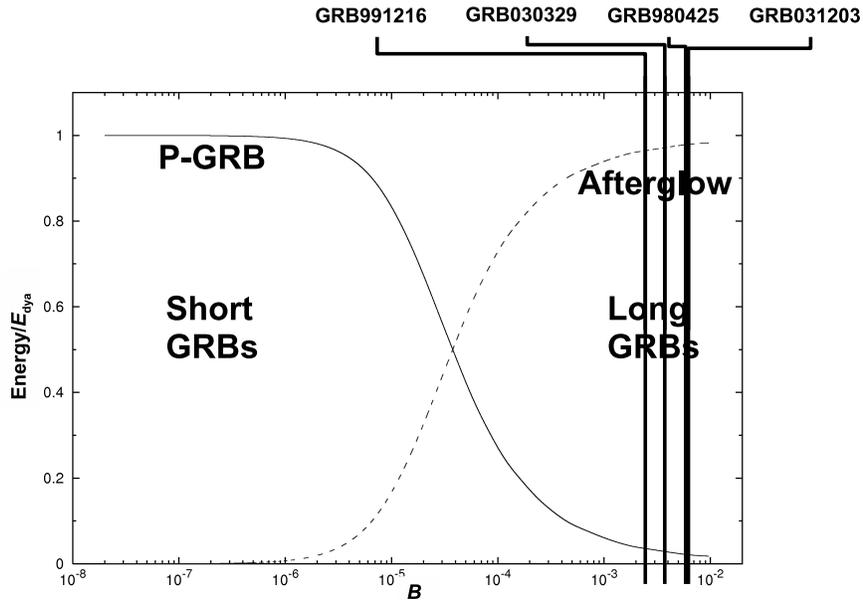}
\caption{The relative intensity of the P-GRB and the afterglow are given for four selected sources as a function of the baryon loading parameter $B$.}
\label{Ruffini2-dia}
\end{figure}

The analysis of the electron positron plasma expansion, far from being purely academical, is essential in describing the entire GRB phenomenon in our model. We are currently working in proving the uniqueness of our model. It is clear that any minimal deviation in the integration of the hydrodynamical equations, or any inadequacy on the integration of the rate equation, may lead to the impossibility of reaching the correct theoretical model of GRBs by lacking the necessary accuracy in the description of the fundamental process determining the energetics of GRBs. Indeed the agreement of our theoretical model \cite{Ruffini2-lett1,Ruffini2-lett2,Ruffini2-lett3,Ruffini2-rbcfx02,Ruffini2-eqts,Ruffini2-eqts2,Ruffini2-Power-Laws} with the observations has been successfully tested in four different sources for the intensities of both the P-GRB and the afterglow in selected energy bands. These sources are: GRB 991216 \cite{Ruffini2-Brasile}, GRB 980425 \cite{Ruffini2-cospar}, GRB 030329 \cite{Ruffini2-030329}, GRB 031203 \cite{Ruffini2-031203} (see Fig. \ref{Ruffini2-dia}). It is particularly interesting that, in all the above sources, the value of the $B$ parameter is smaller than $10^{-2}$ as clearly predicted in Ref. \cite{Ruffini2-rswx00}.

\end{document}